\documentclass[conference]{IEEEtran}
\IEEEoverridecommandlockouts
\usepackage{cite}
\usepackage{amsmath,amssymb,amsfonts}
\usepackage{algorithmic}
\usepackage{graphicx}
\usepackage{textcomp}
\usepackage{xcolor}
\def\BibTeX{{\rm B\kern-.05em{\sc i\kern-.025em b}\kern-.08em
    T\kern-.1667em\lower.7ex\hbox{E}\kern-.125emX}}
\usepackage{siunitx}
\usepackage{booktabs}
\usepackage{hyperref}
\usepackage{tabularx}
\newcolumntype{Y}{>{\centering\arraybackslash}X}

\begin{document}

\title{RadDet: A Wideband Dataset for Real-Time Radar Spectrum Detection}
\author{
    \IEEEauthorblockN{Zi Huang\IEEEauthorrefmark{1}\IEEEauthorrefmark{2}, Simon Denman\IEEEauthorrefmark{1}, Akila Pemasiri\IEEEauthorrefmark{1}, Terrence Martin\IEEEauthorrefmark{1}\IEEEauthorrefmark{2}, Clinton Fookes\IEEEauthorrefmark{1}}
    \IEEEauthorblockA{\IEEEauthorrefmark{1}School of Electrical Engineering \& Robotics, Queensland University of Technology, Brisbane, Australia}
    \IEEEauthorblockA{\IEEEauthorrefmark{2}Revolution Aerospace, Brisbane, Australia}
    Email: \{z37.huang, s.denman, a.thondilege, tl.martin, c.fookes\}@qut.edu.au
}

\maketitle

\begin{abstract}
Real-time detection of radar signals in a wideband radio frequency spectrum is a critical situational assessment function in electronic warfare. Compute-efficient detection models have shown great promise in recent years, providing an opportunity to tackle the spectrum detection problem. However, progress in radar spectrum detection is limited by the scarcity of publicly available wideband radar signal datasets accompanied by corresponding annotations. To address this challenge, we introduce a novel and challenging dataset for radar detection (RadDet), comprising a large corpus of radar signals occupying a wideband spectrum across diverse radar density environments and signal-to-noise ratios (SNR). RadDet contains 40,000 frames, each generated from 1 million in-phase and quadrature (I/Q) samples across a 500 MHz frequency band. RadDet includes 11 classes of radar samples across 6 different SNR settings, 2 radar density environments, and 3 different time-frequency resolutions, with corresponding time-frequency and class annotations. We evaluate the performance of various state-of-the-art real-time detection models on RadDet and a modified radar classification dataset from NIST (NIST-CBRS) to establish a novel benchmark for wideband radar spectrum detection.
\end{abstract}


\section{Introduction}
Real-time detection and classification of complex radar waveforms in degraded signal-to-noise ratio (SNR) environments are fundamental requirements in cognitive electronic warfare (CEW) \cite{robertson2019practical, haigh_cognitive_2021-1}. These functions must be performed at the front-end of a CEW system to detect and localise potential threats  within the radar spectrum. Identifying the dwell time and bandwidth of a radar signal is a critical step in attaining situational awareness of the radar environment, enabling the development of effective radio frequency (RF) countermeasures \cite{robertson2019practical}. Furthermore, in a highly contested electronic warfare (EW) setting with numerous potential RF threats, these tasks must be executed in real-time across a large frequency band to ensure wide spectrum coverage and timely deployment of RF countermeasures \cite{haigh_cognitive_2021-1}. However, the scarcity of publicly available wideband radar signal datasets has limited progress in real-time radar spectrum detection.

In this work, we present a large-scale dataset to support the study of radar spectrum detection in CEW. We focus on the non-cooperative detection of synthetic radar threats across a wide frequency band with overlapping frequency content, and in degraded SNR settings. Using this data, we investigate the application of compute-efficient detection models to perform classification and localisation of time-frequency occupancy, and explore the impact of time-frequency resolution on detection performance and speed. To encourage future research in CEW, we present several dataset variants that explore different synthetic radar density environments. The main contributions of this paper are as follows: (i) we introduce a large-scale open-source dataset\footnote{Our radar spectrum detection datasets can be accessed at: \url{https://github.com/abcxyzi/RadDet}} for wideband radar spectrum detection; and (ii) we establish of a new benchmark for wideband radar spectrum detection across different radar density environments and provide empirical results, analysing the impact of time-frequency resolution on detection performance.

\section{Related Work}
Identifying an RF waveform in the presence of background noise through automatic signal recognition (ASR) has been an active research area over the past decade \cite{scheers2015wideband,schmidt2016kurtosis,oshea_convolutional_2016-1,oshea_unsupervised_2016-1,oshea_over--air_2018,vila2019deep,logue2019expert,huynh-the_automatic_2021}. Recent advancements in ASR have employed various approaches, including the short-time Fourier transform (STFT) \cite{o2017learning}, along with task-specific representations such as the wavelet transform \cite{chen2024augmenting}, Wigner-Ville distribution \cite{huynh2024wavenet}, and Choi-Williams distribution \cite{huynh2021accurate}, to generate time-frequency representations from raw signals for subsequent analysis. The STFT is often preferred due to its consistent time-frequency resolution, while other transforms may introduce undesirable artefacts \cite{scholl2021fourier} in the time-frequency domain and require careful design to be tailored for specific tasks. Other approaches used statistical priors \cite{kakalou2018survey} and feature engineering \cite{vila2019deep,logue2019expert} for feature extraction, however they do not cater for non-cooperative and covert spectrum users encountered in an EW setting. Recent advances in deep convolutional neural networks (CNN) and transformer-based architectures \cite{gong_ast_2021,huang2023multi} have shown promise for ASR, with applications in automatic modulation classification (AMC) \cite{oshea_over--air_2018, jagannath_multi-task_2021-1, huang2023multi}, radar pulse activity segmentation \cite{huang2024multi}, and spectrum sensing \cite{kakalou2018survey,lees2019deep,pemasiri2024automatic}. However, research into radar spectrum detection is limited by the scarcity of publicly available radar detection datasets \cite{geng2021deep}.

The challenge of ASR in an EW setting is the detection and classification of low probability of intercept (LPI) \cite{wan2019lpi,clerico2023lstm} radar signals across a wide frequency band \cite{pace2009detecting,robertson2019practical}. Furthermore, localising the occupancy of radar threats in a wide spectrum within an EW context is particularly difficult due to the need for near real-time performance \cite{haigh_cognitive_2021-1}. While recent advancements in ASR have led to the development of several important radio datasets, such as RadioML \cite{oshea_over--air_2018}, RadarComms \cite{jagannath_multi-task_2021-1,jagannath_multi-task_2022}, DeepRadar \cite{clerico2023lstm}, and RadChar \cite{huang2023multi}, they are constrained by short I/Q sequences, narrow frequency bands, and lack of annotations for time-frequency localisation. RadSeg \cite{huang2024multi} is a large-scale dataset that explores the segmentation of interleaved radar pulse activities while considering long I/Q sequences. However, this dataset focuses on five radar classes and provides temporal annotations only. Another dataset is the RF Dataset of Incumbent Radar Systems in the $3.5$ GHz CBRS Band (NIST-CBRS), which was developed by the National Institute of Standards and Technology (NIST) \cite{caromi2019rf,caromi2021deep}. It provides $40,000$ long-duration waveforms containing $5$ radar classes, sampled at $10$ MHz. Nevertheless, this dataset is limited to the detection of a single radar instance across a narrow SNR range. Other studies \cite{o2017learning,basak2021combined,soltani2022finding} have investigated the near real-time detection of multiple radar and telecommunications waveforms using CNNs, however the RF datasets used in these works are not publicly accessible.

\section{Radar Datasets}
\label{sec:datasets}
We present two large-scale datasets for radar spectrum detection. The first dataset extends the existing work by NIST \cite{caromi2021deep}, providing a new baseline for real-time radar detection. The second dataset, developed in this work, intends to challenge contemporary detection algorithms by exploring a wider radar spectrum, a broader range of radar waveforms, varied radar density environments, and lower SNR settings.

\subsection{NIST CBRS 3.5 GHz Dataset}
\label{ssec:nist}
\begin{figure}[tb]
\centering
\begin{minipage}[b]{0.31\linewidth}
  \centering
  \includegraphics[width=\linewidth]{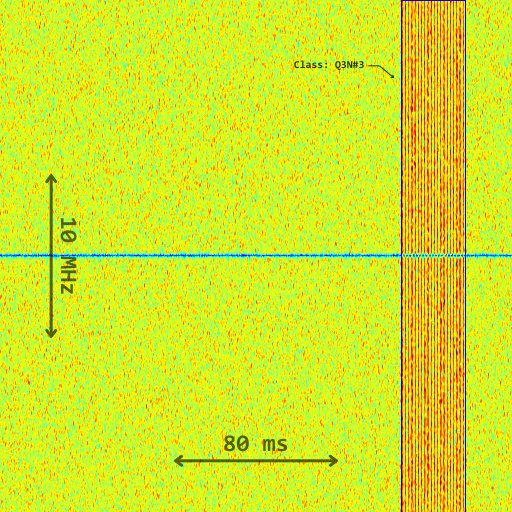}
  \centerline{(a) NIST-CBRS}
\end{minipage}
\hspace{0.01\linewidth} 
\begin{minipage}[b]{0.31\linewidth}
  \centering
  \includegraphics[width=\linewidth]{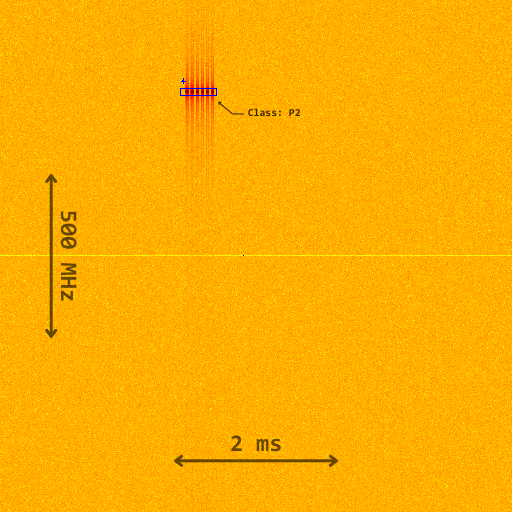}
  \centerline{(b) RadDet-1T}
\end{minipage}
\hspace{0.01\linewidth} 
\begin{minipage}[b]{0.31\linewidth}
  \centering
  \includegraphics[width=\linewidth]{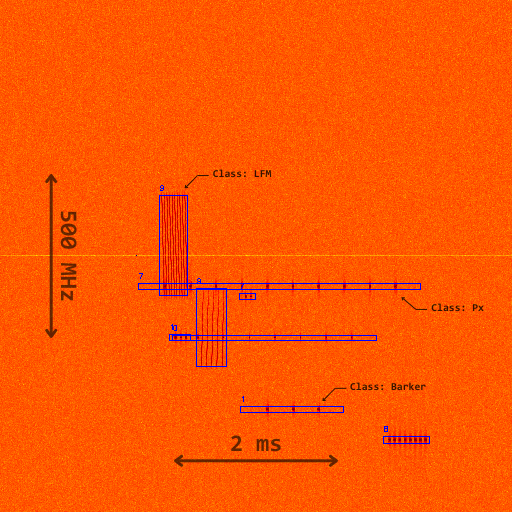}
  \centerline{(c) RadDet-9T}
\end{minipage}

\caption{Radar detection frames with time-frequency annotations (bounding boxes) shown at $10$ \si{\deci\bel} SNR.}
\label{fig:datasets_annotations}
\vspace{-5pt}
\end{figure}

NIST-CBRS \cite{caromi2019rf} is a synthetic radar dataset comprising five radar classes, including two pulsed radars and three linear frequency-modulated (LFM) radars, sampled at $10$ \si{\mega\hertz} with SNR levels ranging from $10$ to $20$ \si{\deci\bel}. The dataset includes $40,000$ baseband I/Q signals, each $800,000$ samples in length. We transform the dataset into a spectrum detection dataset, providing time-frequency annotations as shown in Fig. \ref{fig:datasets_annotations}. The dimensions of the time-frequency bounding box (bbox) for each signal are calculated by 
\begin{equation}
\label{eq:xaxis}
    t_{\text{bbox}} \in \left[ t_s, t_s + \frac{N}{f_\text{prf}} \right],
\end{equation}

\begin{equation}
\label{eq:yaxis}
    f_{\text{bbox}} \in \left[f_{\text{c}} - \frac{B}{2}, f_{\text{c}} + \frac{B}{2}\right],
\end{equation}

\noindent where $t_{\text{bbox}}$ denotes the bbox width (i.e., temporal occupancy), $f_{\text{bbox}}$ denotes the bbox height (i.e., frequency occupancy), $N$ denotes the sample length, $f_\text{c}$ denotes the centre frequency, $f_\text{prf}$ denotes the pulse repetition frequency (PRF), and $B$ denotes the signal bandwidth which is estimated by
\begin{equation}
\label{eq:yaxis_bw}
    B = 
    \begin{cases} 
        B_{\text{chirp}} & \text{if } \text{frequency-modulated}, \\
        t_{\text{pw}}^{-1} + \epsilon & \text{if } \text{pulsed or phase-coded}.
\end{cases}
\end{equation}

For LFM radars (i.e., Q3N classes), $B$ is given by the chirp bandwidth $B_\text{chirp}$. For pulsed radars (i.e., P0N classes), the reciprocal of the signal's pulse width $t_\text{pw}$ is used to estimate $B$. As a clean signal was not provided in the original dataset to precisely compute $B$ for pulsed radars, a constant buffer $\epsilon$ estimated by $0.5 \times t_\text{pw}^{-1}$ is added to account for any measurement uncertainty.

\begin{table}[tb]
\caption{Summary of RadDet Signal Parameters}
\label{table:params}
\small
\def\arraystretch{0.85} 
\noindent\begin{tabularx}{\columnwidth}{>{\raggedright\arraybackslash}X|c|c}
    \toprule
    \textbf{Signal Type} & \textbf{Param.} & \textbf{Range of Value} \\
    \midrule
    All     & $\text{SNR} \, (\si{\deci\bel})$ & $\mathcal{U}(-20, 20, 8)$ \\
            & $f_{\text{c}} \, (\si{\mega\hertz})$ & $\mathcal{U}(-f_{s}/2, f_{s}/2, 10)$ \\
            & $\phi \, (\si{\radian})$ & $\mathcal{U}(-\pi/4, \pi/4, \pi/180)$ \\
            
    \midrule
    Rect (and pulsed)
                & $t_{\text{pw}} \, (\si{\micro\second})$ & $\mathcal{U}(1, 100, 1)$ \\ 
                & $t_{\text{prf}} \, (\si{\kilo\hertz})$ & $\mathcal{U}(10, 50, 1) $ \\
                & $N_{\text{pulse}}$ & $\mathcal{U}(2, 10, 1)$ \\
                
    \midrule
    LFM, FMCW   & $B_{\text{chirp}} \, (\si{\mega\hertz})$ & $\mathcal{U}(10, 100, 1)$ \\
                & $t_{\text{chirp}} \, (\si{\micro\second}$) & $\mathcal{U}(10, 100, 10)$ \\                
                & $N_{\text{chirp}}$ & $\mathcal{U}(1, 20, 1)$ \\
                
    \midrule
    Barker          & $N_{\text{chip}}$ & $\{5, 7, 11, 13\}$ \\
    \midrule
    Frank, Px, P1   & $N_{\text{chip}}$ & $\{4, 9, 16\}$ \\
    \midrule
    P2              & $N_{\text{chip}}$ & $\{4, 16\}$ \\
    \midrule
    P3, P4          & $N_{\text{chip}}$ & $\mathcal{U}(4, 16, 1)$ \\
    \midrule
    Zadoff-Chu      & $N_{\text{chip}}$ & $\{3, 5, 7, 9, 11, 13, 15\}$ \\
    \bottomrule
\end{tabularx}
\end{table}

\subsection{Wideband Radar Detection Dataset}
While NIST-CBRS is valuable as a baseline dataset, it is limited by a narrow $10$ \si{\mega\hertz} frequency band with SNRs confined to $10$ to $20$ \si{\deci\bel}. Critically, it only features at most a single radar instance per frame, which limits its applicability in modelling dense RF scenarios. We introduce RadDet, a novel wideband dataset for radar spectrum detection. RadDet builds upon \cite{huang2024multi} and introduces $11$ radar classes, including $6$ new LPI polyphase codes (P1, P2, P3, P4, Px, Zadoff-Chu) and a new wideband frequency-modulated continuous wave (FMCW), all coexisting across a $500$ \si{\mega\hertz} band. The key parameters used in RadDet are summarised in Table \ref{table:params}. Here, $\mathcal{U}(a, b, c)$ denotes uniform random sampling from $a$ to $b$ with resolution $c$, while $\{ \cdot \}$ denotes uniform random sampling from a discrete set. $N_{i}$ denotes the number of parameters given by $i$. 

Unlike existing RF classification datasets \cite{oshea_over--air_2018,jagannath_multi-task_2021-1,huang2023multi}, RadDet provides long sequences, each consisting of $1$ million complex baseband I/Q samples ($\Vec{x}_{\text{i}} + j \Vec{x}_{\text{q}}$). We generate a total of $40,000$ radar frames and distribute the dataset in three parts. The training set contains $20,000$ signals, while the validation and test sets contain $14,000$ and $6,000$ signals, respectively. Additive white Gaussian noise (AWGN) is added to each signal to simulate varying SNR settings. We sample SNR from a uniform distribution to produce signal frames that fall within $-20$ and $20$ \si{\deci\bel} at a resolution of $8$ \si{\deci\bel}. This means that there are more than 6,500 unique signals genreated per SNR. The data splits are carefully chosen to maintain a similar inter-class distribution to \cite{caromi2019rf} across each SNR.

\begin{figure}[tb]
\centering
\begin{minipage}[b]{0.31\linewidth}
  \centering
  \includegraphics[width=\linewidth]{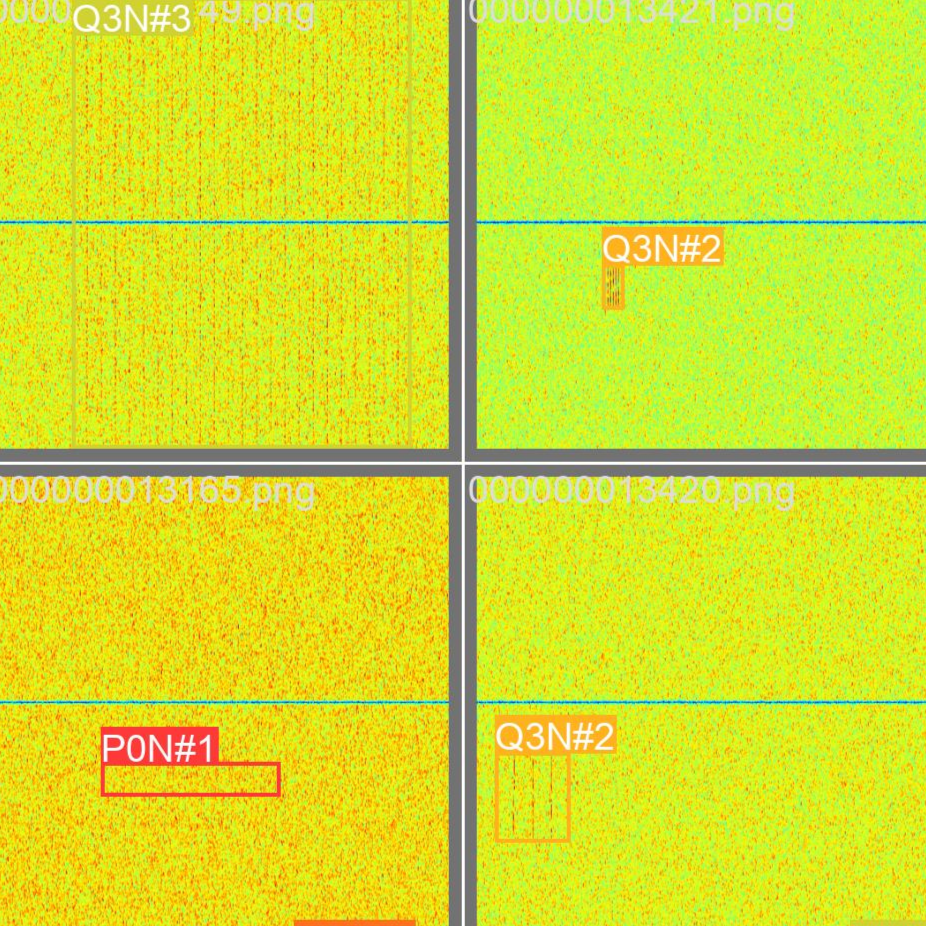}
  \centerline{(a) NIST-CBRS}
\end{minipage}
\hspace{0.01\linewidth} 
\begin{minipage}[b]{0.31\linewidth}
  \centering
  \includegraphics[width=\linewidth]{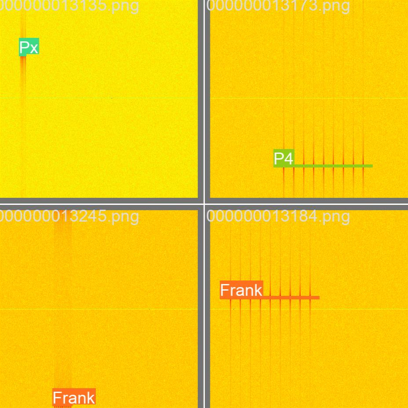}
  \centerline{(b) RadDet-1T}
\end{minipage}
\hspace{0.01\linewidth} 
\begin{minipage}[b]{0.31\linewidth}
  \centering
  \includegraphics[width=\linewidth]{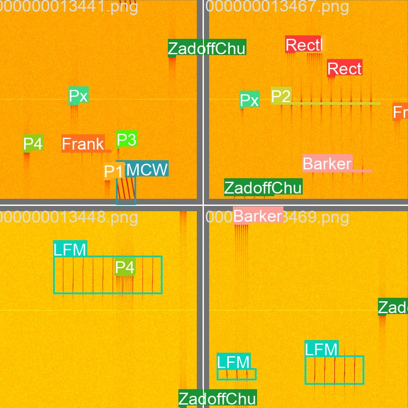}
  \centerline{(c) RadDet-9T}
\end{minipage}

\vspace{5pt}

\begin{minipage}[b]{0.31\linewidth}
  \centering
  \includegraphics[width=\linewidth]{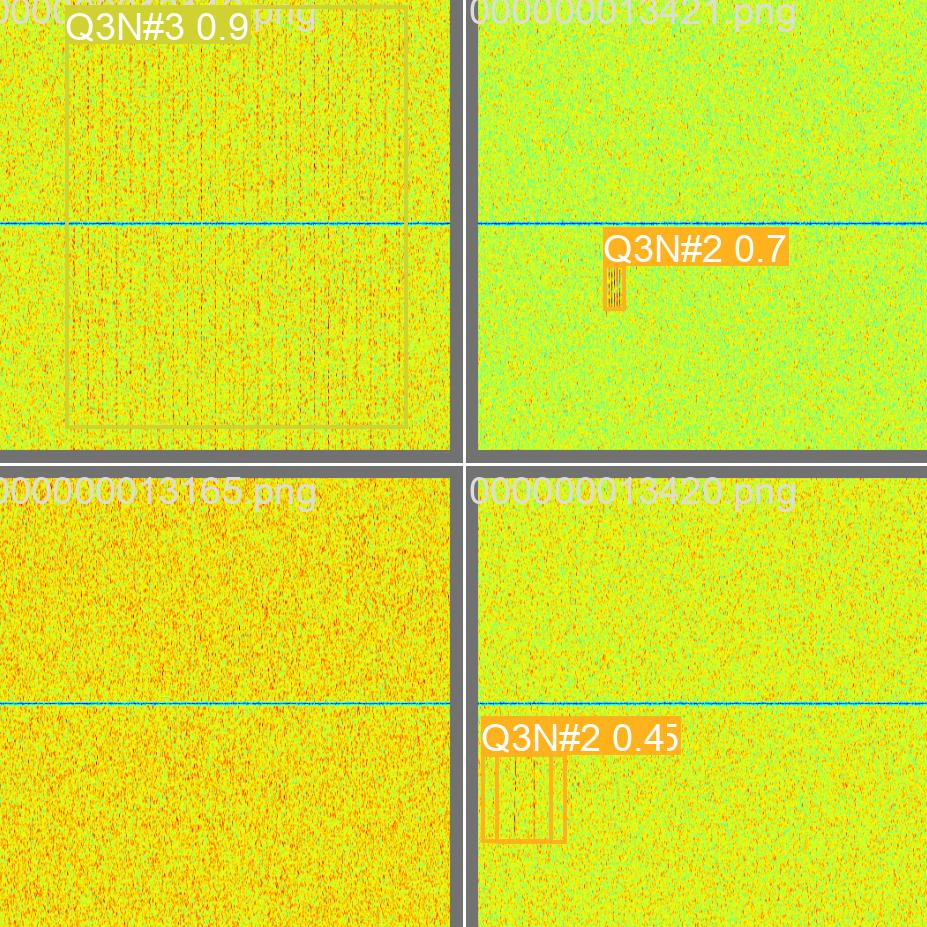}
  \centerline{(d) NIST-CBRS}
\end{minipage}
\hspace{0.01\linewidth} 
\begin{minipage}[b]{0.31\linewidth}
  \centering
  \includegraphics[width=\linewidth]{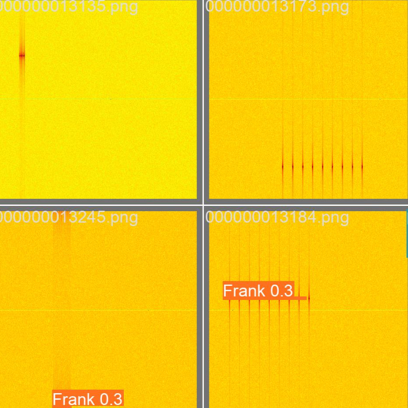}
  \centerline{(e) RadDet-1T}
\end{minipage}
\hspace{0.01\linewidth} 
\begin{minipage}[b]{0.31\linewidth}
  \centering
  \includegraphics[width=\linewidth]{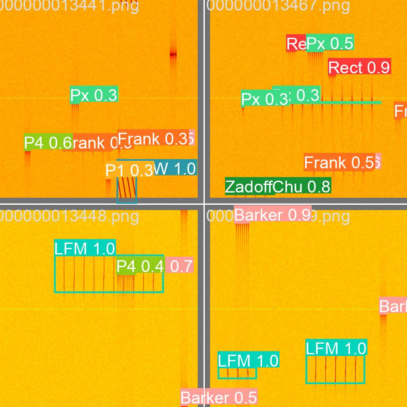}
  \centerline{(f) RadDet-9T}
\end{minipage}

\caption{Radar detection results at $10$ \si{\deci\bel} SNR. Ground truth frames in (a)–(c) correspond to model predictions in (d)–(f), respectively. The confidence scores of each bounding box prediction are also shown in (d)–(f).} \label{fig:inference_res}
\vspace{-5pt}
\end{figure}

\begin{table}[tb]
\caption{Radar Detection Dataset Variants}
\label{table:compressions}
\small 
\def\arraystretch{0.75} 
\noindent\begin{tabularx}{\columnwidth}{>{\raggedright\arraybackslash}X|c|c|c|c}
    \toprule
    \textbf{Dataset}  & \textbf{Size} & \textbf{$N_{\text{pool}}$} & \textbf{Dim.} & \textbf{Res. (\si{\micro\second}, \si{\mega\hertz})} \\
    \midrule
    NIST-CBRS       & $\text{S}$ & $60$ & $128^{2}$     & $624, \quad 0.08$ \\
    ($f_{s} = 10~\si{\mega\hertz}$)     & $\text{M}$ & $30$ & $256^{2}$     & $313,  \quad 0.04$ \\
     & $\text{L}$ & $15$ & $512^{2}$     & $157, \quad 0.02$ \\
    \midrule
    RadDet-9T/1T    & $\text{S}$ & $75$ & $128^{2}$  & $15.6,  \quad 3.91$ \\
    ($f_{s} = 500~\si{\mega\hertz}$) & $\text{M}$ & $19$ & $256^{2}$  & $7.81,  \quad 1.95$ \\
     & $\text{L}$ & $4$ & $512^{2}$   & $3.91,  \quad 0.98$ \\
    \bottomrule
\end{tabularx}
\end{table}

To investigate wideband spectrum detection in different scenarios, we provide RadDet in two different radar environments: dense and sparse. Similar to NIST-CBRS, our sparse dataset (RadDet-1T) provides at most a single radar instance per frame whereby the probability of a radar being present in a scene is $50$\%. Our dense dataset (RadDet-9T) contains up to $9$ radar instances per frame where the probability of background (noise-only) frames is $10$\%. RadDet-9T represents a conservative dense maritime radar environment, as described in \cite{robertson2019practical}, where up to $200$ radars may be observed within a typical $2$-second surveillance period. We consider uniform random sampling to independently generate different permutations of radar encounters and signal attributes for each frame. This means that radar activities in each frame are uniformly distributed across a fixed time horizon ($2$ \si{\milli\second}) and a fixed frequency band ($500$ \si{\mega\hertz}). To ensure that the signal parameters chosen for each radar class reflect typical radars  encountered in EW, we adopt the emitter parameter library from \cite{alma991010191960904001} to bound our sampling range. For example, we consider large sweep bandwidths of up to $100$ \si{\mega\hertz} for LFM signals, whereas the bandwidths of LFM radars are typically much lower outside the context of EW \cite{pace2009detecting}. We transform the baseband I/Q signals into spectrograms using the STFT. The approach in \cite{caromi2021deep} is adapted to generate max-hold spectrograms as a compute-efficient way to represent wideband signals. Due to the large spread of signal parameters considered in RadDet, no window functions are applied to avoid introducing pre-processing bias towards certain classes. The size of each max-hold spectrogram is calculated by
\begin{equation}
\label{eq:dims_t}
    \dim(t) = \frac{N - N_{\text{overlap}}}{N_{\text{pool}}(N_{\text{seg}} - N_{\text{overlap}})},
\end{equation}

\begin{equation}
\label{eq:dims_f}
    \dim(f) = N_{\text{nfft}},
\end{equation}

\noindent where $N_{\text{nfft}}$ denotes the number of frequency bins computed by the STFT, $N_{\text{seg}}$ denotes length of the time segment used for computing the STFT, $N_{\text{overlap}}$ denotes the overlap between consecutive time segments, and $N_{\text{pool}}$ represents the number of time bins used in the max-hold operation which controls the level of compression applied. While the max-hold operation reduces temporal resolution, it is necessary to enable efficient processing of extremely long I/Q sequences. The degree of compression applied should be scenario-dependent in an EW context, and balance the trade-off between detection accuracy and speed. We provide the RadDet dataset as spectrograms in $3$ resolutions as shown in Table \ref{table:compressions} where bbox annotations are computed per (\ref{eq:xaxis}) and (\ref{eq:yaxis}). Example of resultant annotations are depicted in Fig. \ref{fig:datasets_annotations}(b) and Fig. \ref{fig:datasets_annotations}(c). Annotations are normalised and structured using the YOLO format \cite{redmon2016you}.

\begin{table*}[tb]
\caption{Performance of Detection Models on NIST-CBRS and RadDet-9T/1T Test Sets}
\label{table:massive_table}
\small
\def\arraystretch{0.75}
\begin{tabularx}{\textwidth}{>{\raggedright\arraybackslash}p{1.7cm}@{\hspace{1em}}|c|Y|Y|Y|Y|Y|Y|Y|Y}
    \toprule
    \textbf{Model} & \textbf{Size} & \textbf{\#Param.} & \multicolumn{2}{c|}{\textbf{NIST-CBRS ($\%$)}} & \multicolumn{2}{c|}{\textbf{RadDet-9T} ($\%$)} & \multicolumn{2}{c|}{\textbf{RadDet-1T} ($\%$)} & $\textbf{FPS}_{\text{avg}}$ \\
    & & ($\text{M}$) & \textbf{$\text{mAP}_{50}$} & \textbf{$\text{mAP}_{50:95}$} & \textbf{$\text{mAP}_{50}$} & \textbf{$\text{mAP}_{50:95}$} & \textbf{$\text{mAP}_{50}$} & \textbf{$\text{mAP}_{50:95}$} & \\
    \midrule
    
    YOLOv3 & $\text{S}$ & $103.67$ & $87.20$ & $67.28$  & $25.68$ & $19.10$ & $19.62$ & $13.28$& ${1023.0}$ \\
    YOLOv6-M & & $51.98$ & $86.57$ & $66.37$ & $23.02$ & $16.49$ & $10.41$ & $4.93$ & $\mathbf{1301.8}$ \\
    YOLOv9-M & & $20.02$ & $85.88$ & $67.72$ & $\mathbf{25.93}$ & $\mathbf{19.47}$ & $\mathbf{20.10}$ & $\mathbf{13.29}$ & $896.6$ \\
    RT-DETR-L & & $31.99$ & $\mathbf{91.27}$ & $\mathbf{73.29}$ & $24.70$ & $17.77$ & $18.75$ & $12.52$ & $228.5$ \\

    \midrule
    YOLOv3 & $\text{M}$ & $103.67$ & $93.52$ & $72.96$ & $\mathbf{35.69}$ & $\mathbf{29.82}$ & $24.81$ & $18.67$ & $688.2$ \\
    YOLOv6-M & & $51.98$ & $92.97$ & $75.86$ & $31.45$ & $25.96$ & $22.68$ & $16.49$ & $\mathbf{921.3}$ \\
    YOLOv9-M & & $20.02$ & $93.81$ & $\mathbf{79.12}$ & $35.10$ & $29.14$ & $\mathbf{26.03}$ & $\mathbf{19.73}$ & $579.2$ \\
    RT-DETR-L & & $31.99$ & $\mathbf{94.69}$ & $78.70$ & $26.20$ & $19.43$ & $20.07$ & $14.55$ & $210.5$ \\

    \midrule
    YOLOv3 & $\text{L}$ & $103.67$ & $93.70$ & $72.62$ & $\mathbf{60.37}$ & $\mathbf{53.97}$ & $\mathbf{31.85}$ & $\mathbf{25.41}$ & $274.2$ \\
    YOLOv6-M & & $51.98$ & $88.23$ & $70.82$ & $43.57$ & $38.14$ & $24.66$ & $18.18$ & $\mathbf{470.9}$ \\
    YOLOv9-M & & $20.02$ & $91.31$ & $72.97$ & $47.37$ & $40.85$ & $31.41$ & $24.56$  & $449.4$ \\
    RT-DETR-L & & $31.99$ & $\mathbf{95.31}$ & $\mathbf{80.96}$ & $29.34$ & $19.90$ & $20.90$ & $15.25$ & $177.7$ \\
    \bottomrule
\end{tabularx}
\end{table*}

\section{Experiments}

\subsection{Detection Models}
We establish a performance baseline for wideband radar spectrum detection by considering real-time detection models, all targeting multi-class radar detection on both the NIST-CBRS and RadDet datasets. We consider real-time object detectors that are competitive in computer vision tasks, including YOLOv3 \cite{redmon2018yolov3}, YOLOv6 \cite{li2022yolov6}, YOLOv9 \cite{wang2024yolov9}, and RT-DETR \cite{zhao2024detrs}. These detectors provide a comprehensive representation of baseline performance, allowing us to investigate the trade-off between detection performance and inference speed across different input resolutions and radar environments.

We train and evaluate each model on a single Nvidia Tesla A100 GPU. The models are trained for $100$ epochs using the AdamW optimiser with a linear learning rate scheduler, where the learning rate is bounded between $0.01$ and $0.001$, and a warm-up period of $3$ epochs at $0.1$. To improve generalisation and training stability, we normalise spectrogram inputs to the range $[0,1]$ and consider contemporary data augmentation techniques, including translations ($10\%$), scaling ($50\%$), erasing ($20\%$), mixup ($20\%$), mosaic ($2\times2$), and copy-paste ($20\%$) \cite{ghiasi2021simple}. We also train models without data augmentation to serve as a comparison. All models are evaluated on test sets where the best models are selected based on their best validation performance during training.

\subsection{Radar Detection Benchmark}
We evaluate the baseline performance of detection models suitable for real-time applications on the NIST-CBRS and RadDet-9T/1T test sets through the following experiments: (i) comparison of baseline architectures; (ii) impact of time-frequency resolution on detection accuracy and speed; and (iii) changes in detection performance across different radar density environments and SNR settings. Performance is measured using inference speed, represented by the average frame rate ($\text{FPS}_{\text{avg}}$), and mean average precision ($\text{mAP}$) reported at two intersection-over-union (IoU) \cite{everingham2010pascal} settings: $50\%$ ($\text{mAP}_{50}$) and $50$ to $95\%$ at $5\%$ increments ($\text{mAP}_{50:95}$).

\begin{figure}[tb]
\begin{minipage}[b]{0.49\linewidth}
  \centering
  \centerline{\includegraphics[width=1\linewidth]{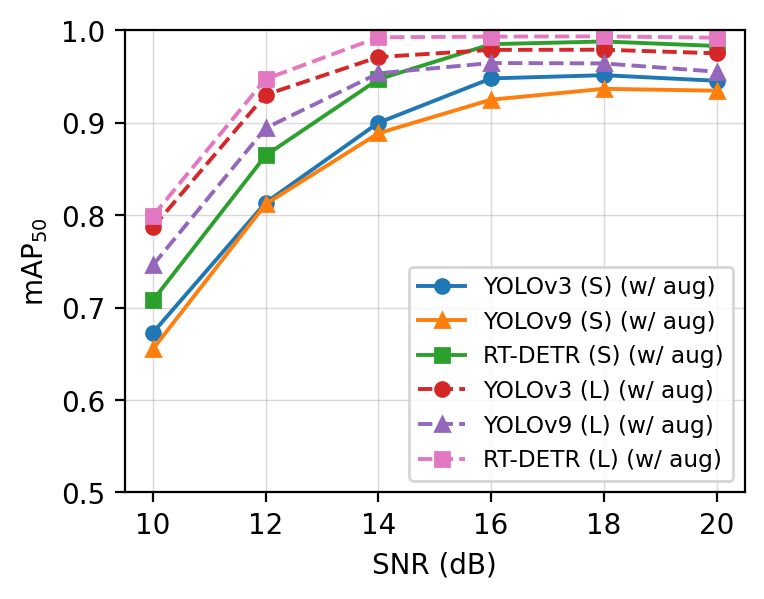}}
  \centerline{(a) NIST-CBRS}\medskip
\end{minipage}
\hfill
\begin{minipage}[b]{0.49\linewidth}
  \centering
  \centerline{\includegraphics[width=1\linewidth]{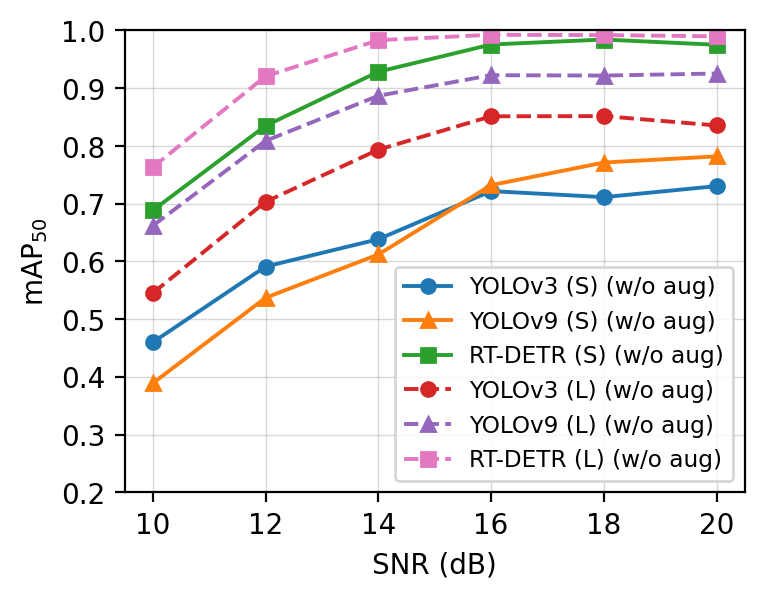}}
  \centerline{(b) NIST-CBRS (w/o aug)}\medskip
\end{minipage}
\hfill
\begin{minipage}[b]{0.49\linewidth}
  \centering
  \centerline{\includegraphics[width=1\linewidth]{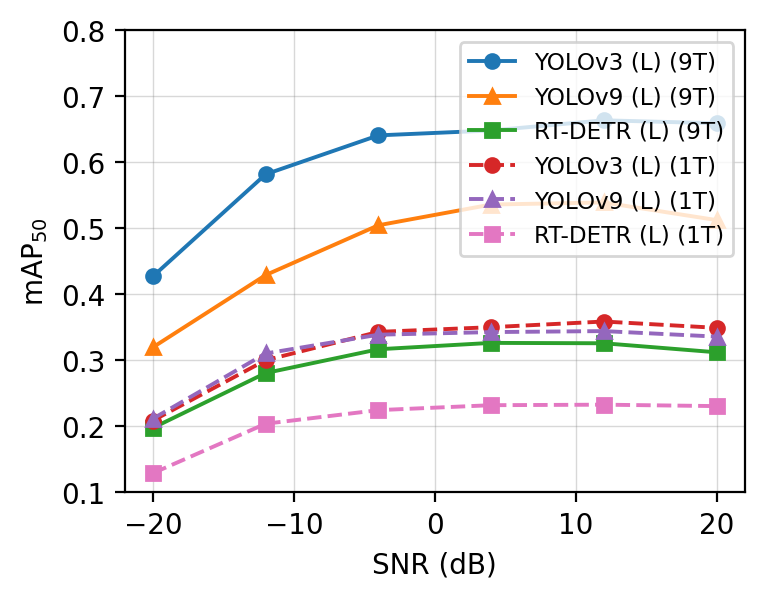}}
  \centerline{(c) RadDet-9T/1T}\medskip
\end{minipage}
\hfill
\begin{minipage}[b]{0.49\linewidth}
  \centering
  \centerline{\includegraphics[width=1\linewidth]{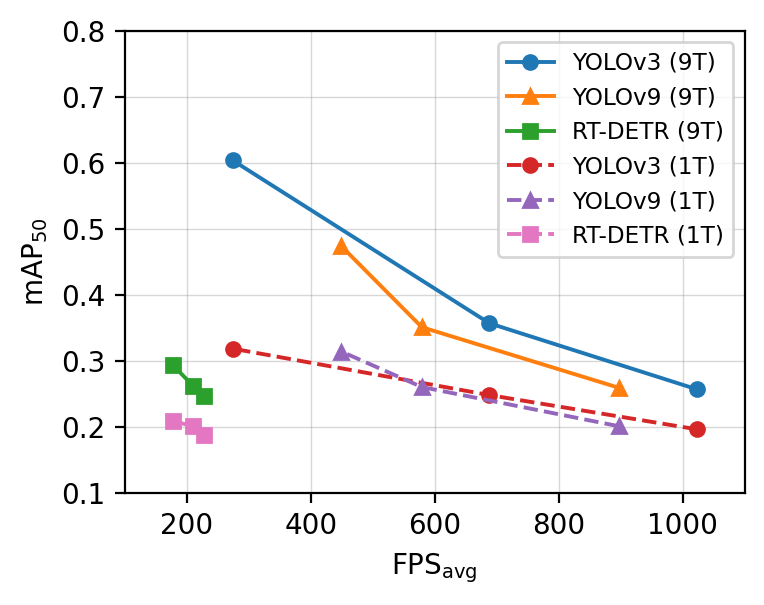}}
  \centerline{(d) $\text{mAP}_{50}$ vs $\text{FPS}_{\text{avg}}$}\medskip
\end{minipage}
\hfill
\vspace{-20pt} 
\caption{Comparison of detection performance of models evaluated on NIST-CBRS and RadDet-9T/1T across different SNR settings. Unless otherwise noted, results reflect the use of data augmentation during training.}
\label{fig:nice_stack_of_plots}
\vspace{-5pt}
\end{figure}

Table \ref{table:massive_table} presents a summary of results. The metrics $\text{mAP}_{50}$ and $\text{mAP}_{50:95}$ presented in Table \ref{table:massive_table} are mean values computed across all $6$ SNR settings from the respective datasets. Several interesting observations can be made. First, data augmentation greatly enhances detection performance across all models without impacting inference speed, as shown in Fig. \ref{fig:nice_stack_of_plots}(a)–(b). Second, there is a distinct trade-off between $\text{mAP}$ and $\text{FPS}_{\text{avg}}$ across all models. Higher time-frequency resolution inputs provide better $\text{mAP}$ but at the cost of lower $\text{FPS}_{\text{avg}}$ due to increased computational demands. YOLOv3 at an input size of $128^{2}$ (S) is $49\%$ faster than at $256^{2}$ (M), with a corresponding $\text{mAP}_{50}$ reduction of $7\%$ and $39\%$ on NIST-CBRS and RadDet-9T, respectively. This trend is highlighted in Fig. \ref{fig:nice_stack_of_plots}(d). Third, the YOLO family of models outperforms RT-DETR on RadDet-9T, which is expected as the detection of small objects is a known limitation \cite{zhao2024detrs} of RT-DETR. Furthermore, RT-DETR relies on pre-trained embeddings from natural images, making it less suited for radar spectrum detection. 

Lastly, the sparse radar environment in RadDet-1T poses challenges for all models across all SNR settings, as shown in Fig. \ref{fig:inference_res}. This challenge stems from two key factors: the presence of small objects (e.g., LPI radar classes) and the infrequent occurrence of radar pulses in a sparse signal environment. In contrast, radar classes in NIST-CBRS typically occupy larger bounding box footprints due to the lower spectrum bandwidth, providing a higher ratio of useful signal pixels to noise pixels for detection models, leading to better detection performance (Fig. \ref{fig:nice_stack_of_plots}). These characteristics of RadDet make it a challenging dataset for real-time detection models.

\section{Conclusion}
We have introduced RadDet, a new open-source dataset designed for the challenging task of wideband radar spectrum detection. This dataset provides $11$ challenging classes of radar samples across $6$ different SNR settings, $2$ radar density environments, and $3$ different time-frequency resolutions with corresponding time-frequency annotations. Our baseline results demonstrate the feasibility of applying real-time detectors for wideband radar spectrum detection, establishing a first-of-its-kind benchmark that explores the trade-off between time-frequency resolution, detection accuracy, and inference speed for real-time radar spectrum detection. 

\clearpage

\bibliographystyle{IEEEbib}
\bibliography{refs}

\end{document}